\documentclass[groupedaddress,superscriptaddress,
reprint,
amsmath,amssymb,
aps,
colorlinks=true, breaklinks=true, linkcolor=darkblue, citecolor=darkblue, urlcolor = darkblue
]{revtex4-2}
\usepackage{cancel}
\usepackage{graphicx}% Include figure files
\usepackage{dcolumn}% Align table columns on decimal point
\usepackage{bm}% bold math

\usepackage{siunitx}
\usepackage{orcidlink}

\usepackage{times}

\usepackage{xcolor}
\definecolor{darkblue}{rgb}{0, 0, 0.8}
\usepackage{lipsum}
\newcommand{\PASQAL}{PASQAL SAS, 7 Rue Léonard de Vinci, 91300 Massy, France}
\newcommand{\IOGS}{Université Paris-Saclay, Institut d’Optique Graduate School,
CNRS, Laboratoire Charles Fabry, 91127 Palaiseau Cedex, France}

\begin{document}

%\preprint{APS/123-QED}

\title{Rearrangement of single atoms in a 2000-site optical tweezers array at cryogenic temperatures}

\author{Grégoire Pichard}
\affiliation{\PASQAL}\affiliation{\IOGS}
\author{Desiree Lim}
\affiliation{\PASQAL} 
\author{\'Etienne Bloch} 
\affiliation{\PASQAL}
\author{Julien Vaneecloo}
\affiliation{\PASQAL}
\author{Lilian Bourachot} 
\affiliation{\PASQAL}
\author{Gert-Jan Both}
\affiliation{\PASQAL}
\author{Guillaume Mériaux}
\affiliation{\PASQAL}
\author{Sylvain Dutartre}
\affiliation{\PASQAL}
\author{Richard Hostein}
\affiliation{\PASQAL}
\author{Julien Paris}
\affiliation{\PASQAL}
\author{Bruno Ximenez}
\affiliation{\PASQAL}
\author{Adrien Signoles}
\affiliation{\PASQAL}
\author{Antoine Browaeys}
\affiliation{\IOGS}
\author{Thierry Lahaye}
\affiliation{\IOGS}
\author{Davide Dreon}
\affiliation{\PASQAL}
\date{\today}

\begin{abstract}
We report on the trapping of single rubidium atoms in large arrays of optical tweezers comprising up to 2088 sites in a cryogenic environment at $6\,\mathrm{K}$. Our approach relies on the use of microscope objectives that are \text{in-vacuum} but at room temperature, in combination with windowless thermal shields into which the objectives are protruding to ensure a cryogenic environment for the trapped atoms. To achieve enough optical power for efficient trapping, we combine two lasers at slightly different wavelengths. We discuss the performance and limitations of our design. Finally, we demonstrate atom-by-atom rearrangement of an 828-atom target array using moving optical tweezers controlled by a field-programmable gate array.
\end{abstract}

\maketitle

Arrays of cold neutral atoms constitute an excellent platform to achieve microscopic control of single qubits. In recent years, there has been a number of impressive results, using such a platform for analog~\cite{browaeys2020many,ebadi2021quantum,scholl2021quantum} and digital quantum computing~\cite{weiss2017quantum,henriet2020quantum,kaufman2021quantum,Bluvstein2024}. So far, this has been achieved on arrays of up to a few hundreds of atoms. Extending these results at the scale of thousands of atoms is currently the subject of a major research effort~\cite{huft2022,pause2024supercharged,norcia2024iterative,gyger2024continuous}, with a recent breakthrough at the scale of more than 6,000 atoms~\cite{manetsch2024tweezer}. In addition, using a cryogenic environment for the atoms would come with the advantage of a better vacuum~\cite{schymik2021single} and a longer Rydberg state lifetime, especially for the case of circular states~\cite{cantat2020long, cohen2021quantum, holzl2024long}. These two advantages directly result in improved register preparation and better fidelity for quantum operations. 

However, combining large tweezers arrays with a cryogenic environment remains technically challenging. Firstly, the high laser power required to generate a large-scale tweezers array can result in too strong a thermal load for the cryostat. Secondly, combining high-numerical-aperture, large-field-of-view optics (typically, a refractive microscope objective) together with a cryogenic setup is nontrivial. In this work, we investigate a possible approach to address those challenges.

\begin{figure*}[tbh!]
\centering
\includegraphics[]{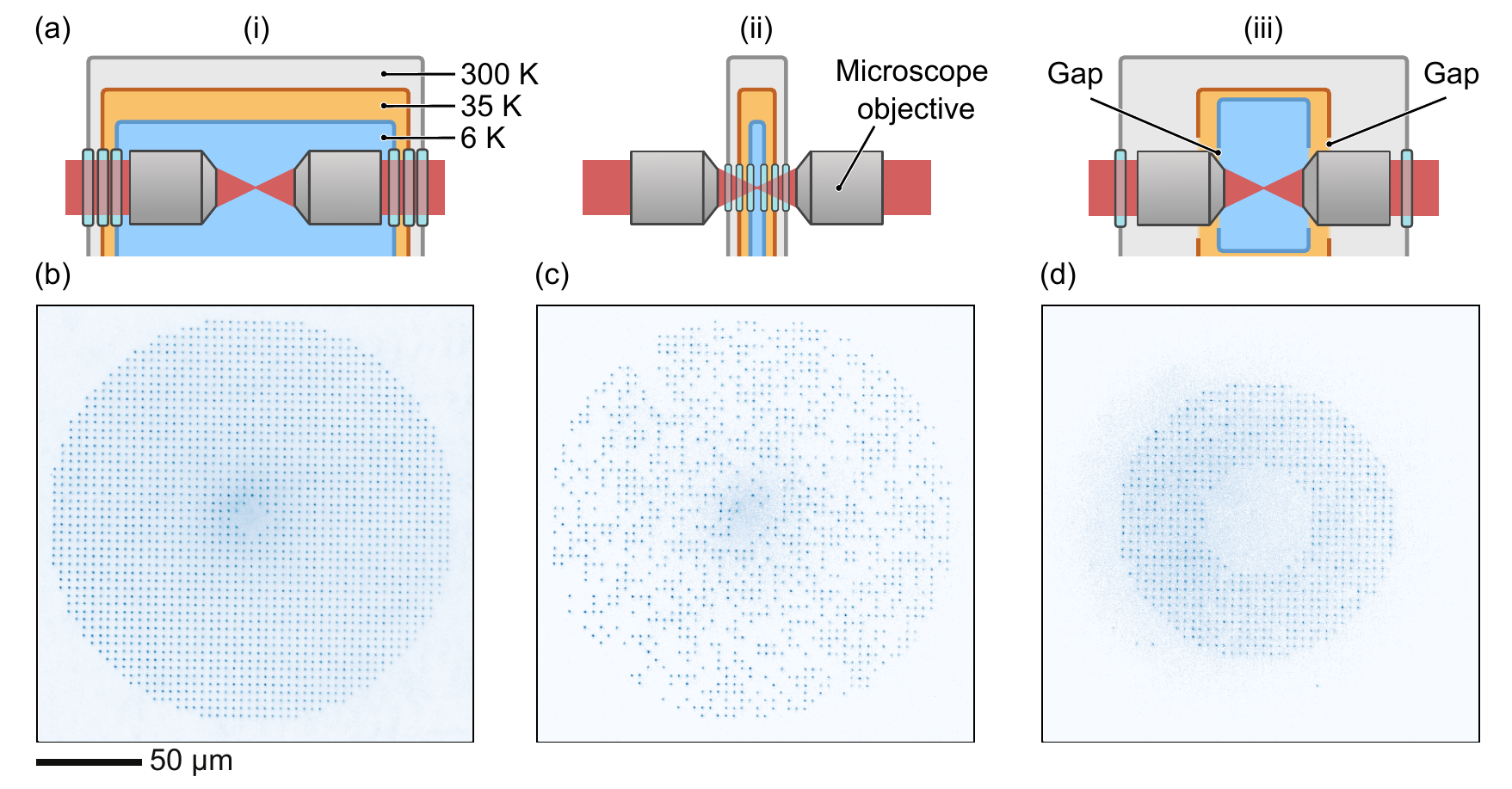}
\caption{(a) Three possible ways to combine the use of microscope objectives and a cryogenic setup with (i) the objectives inside the cryogenic environment; with (ii) the objectives outside the vacuum chamber, or with (iii) the objectives under vacuum but at room temperature. The latter is the solution explored in this work. (b) Averaged fluorescence image of single atoms trapped in a 2088-trap array.  (c) Single-shot fluorescence image in the same array, showing 1103 trapped atoms. (d) Rearrangement with a target array of 828 atoms selected from a 1824-trap array; the final array shows a $\sim95\%$ occupancy.}
\label{fig1}
\end{figure*}

Figure~\ref{fig1}~(a) shows three possible ways to combine a microscope objective with a cryogenic setup. A first, natural approach would be to embed the optics in the low-temperature environment (Fig.~\ref{fig1}~(a), (i)), as we did in our previous work~\cite{schymik2021single} for aspherical lenses; however the design of cryogenic microscope objectives with custom specifications is notoriously difficult. In addition, for such a configuration the losses and reflections inside the objective lead to unacceptable heat loads for laser powers beyond a couple of watts, putting a strong limit on the number of achievable traps. A second approach would be the extreme opposite, where the objectives are outside the vacuum chamber (Fig.~\ref{fig1}~(a), (ii)). However, this comes with a number of complications: indeed, the optical design needs to carefully account for many windows, that are at different temperatures, and thus whose positions are hard to control with high accuracy due to thermal contraction. Moreover, keeping a reasonable working distance for the objectives imposes stringent constraints on the volume of the vacuum chamber. In this work, we therefore explore the configuration depicted in Fig.\ref{fig1}~(a),~(iii) where custom, ultra-high-vacuum-compatible objectives at room temperature cut into \emph{windowless} thermal shields. This might seem a risky approach, as the unavoidable gaps in the shields around the objectives may compromise both the achievement of cryogenic temperatures (we will see below that this is not case) and the quality of the vacuum (this, however, turns out to be the case). With this approach we nevertheless achieve efficient trapping of single atoms in arrays with up to 2088 sites in an environment with a base temperature of $6\,\mathrm{K}$, and demonstrate the  rearrangement of arrays of up to 828 traps with very few defects. We discuss the technical details of our design, characterize the performances of the setup and explore possible future improvements.

\begin{figure*}[tbh!]
\centering
\includegraphics[width=1\textwidth]{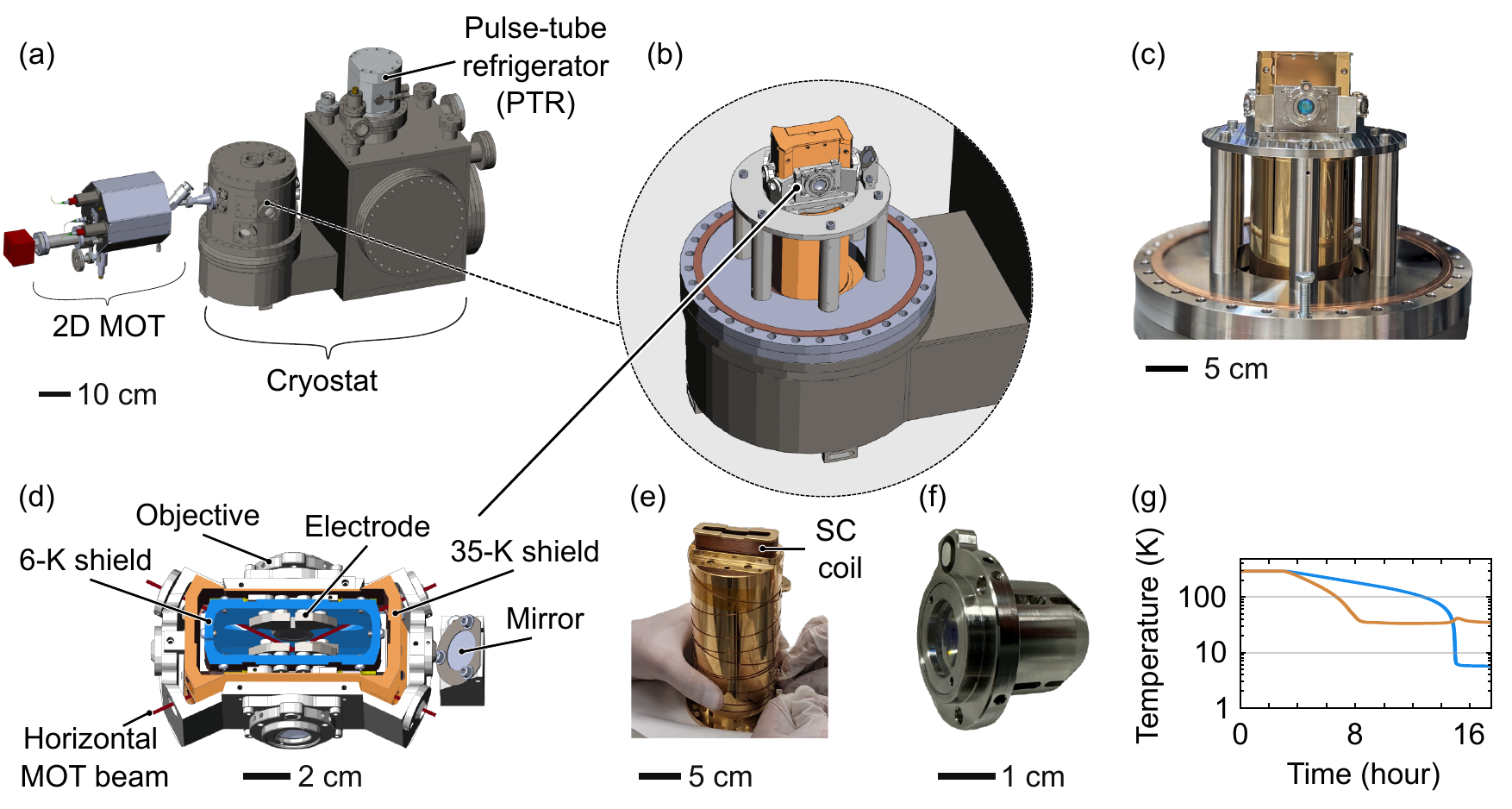}
\caption{(a) Diagram of the whole setup including a 2D MOT as the atomic source and the cryostat. The interior of the science chamber is detailed in the magnified view (b). (c) Photograph of the objectives mounted on a $300\,\mathrm{K}$ support that surrounds the gold-plated copper shield at $35\,\mathrm{K}$. (d) Transverse cross section showing how the objectives are arranged with the two shields at $35\,\mathrm{K}$ and $6\,\mathrm{K}$.  (e) Photograph illustrating the lower, elongated superconducting (SC) coil for the MOT gradient; it is fixed onto the $6\,\mathrm{K}$ shield. (f) Close-up shot of one of our custom, UHV-compatible objectives. (g) Temperature behavior of the first (brown) and second stage (blue) during cool-down.}
\label{fig2}
\end{figure*}

We first summarize in Fig.~\ref{fig1}~(b-d) the main achievements of this work, by showing fluorescence images of single $^{87}$Rb atoms trapped in large optical tweezers arrays. Figure~\ref{fig1}~(b) is an averaged fluorescence image~\footnote{In practice, in order to increase the contrast of this occupation-averaged image, we use the following procedure: we accumulate images with $20\,\mathrm{ms}$ integration time for $\sim 2$~s, and then display (independently for each pixel), the maximum fluorescence level obtained over the whole series of images.} of a circular patch of a square lattice of $2088$ optical tweezers with a 3.7~µm nearest-neighbor spacing; the central hole in the array arises from blocking the zeroth-order beam from the spatial light-modulator used to create the array (see below). Efficient single-atom loading is achieved even for traps on the outside border of the array, that lie at almost 100~µm from the optical axis. This illustrates in a striking manner the advantage of using microscope objectives, as compared to aspherical lenses~\cite{schymik2022situ}, for the generation of large arrays. Figure~\ref{fig1}~(c) shows a single shot image of atoms loaded in this array, containing 1103 trapped atoms, i.e. $\simeq53\%$ trap occupancy. Finally, in Fig.~\ref{fig1}~(e) we show the result of a rearrangement procedure, using sequential atom moves with a single moving optical tweezers as in~\cite{barredo2016atom}. Within an array of 1824 traps, we select a target sub-array of 828 traps. The obtained occupancy of the rearranged array is about $95\%$, limited by the finite trapping lifetime ($\sim 100$~s) and by imperfections in the rearrangement (see more details below).

We now describe in details the experimental setup, which is a significant evolution over our previous work~\cite{schymik2021single}. The overall apparatus is depicted in figure~\ref{fig2}~(a). We chose to keep the same custom-made cryostat, based on a large, unbaked ultra-high vacuum chamber accommodating a pulse-tube refrigerator (PTR). The atomic source is now replaced by a 2D MOT which is more compact than the Zeeman slower used in \cite{schymik2021single}. The two-stage PTR is used to cool down copper thermal shields (that have been polished and gold-plated to reduce their emissivity) at $\sim 35\,\mathrm{K}$ and  $\sim 6\,\mathrm{K}$. Two ion pumps with non-evaporable getter cartridges, not pictured on Fig.~\ref{fig2}, are also added to pump the science chamber.

Figure~\ref{fig2}~(b) shows the inside of the science chamber, where ultra-high vacuum (UHV) objectives [Fig.~\ref{fig2}~(f)], that are pre-aligned at room temperature, are mounted on a $300\,\mathrm{K}$ support. These custom-made objectives (manufactured by Special Optics) offer many advantages over the aspherical lenses used in~\cite{schymik2021single}: a larger field of view of $\pm130\,\mathrm{\mu m}$, a higher numerical aperture of $0.57$, and chromatic correction (over the range of wavelengths between $760\,\mathrm{nm}$ and $850\,\mathrm{nm}$, the chromatic shift is below $1\,\mathrm{\mu m}$). This makes it much more convenient to align the fluorescence imaging system and enables us to use different wavelengths for the trap laser with the same alignment. Furthermore, a small mirror, mounted above the entrance of the pupil and  referenced to the optical axis, eases beam alignment. The front surfaces of the objectives are coated with a $100\,\mathrm{nm}$ indium-tin oxyde (ITO) coating to avoid stray electric fields in view of future experiments involving Rydberg states. The working distance of these objectives is $7.25\,\mathrm{mm}$ and they have a transmission of $85\%$ (per objective) within the range $760$ - $850\,\mathrm{nm}$.

\begin{figure*}[tb!]
\centering
\includegraphics[]{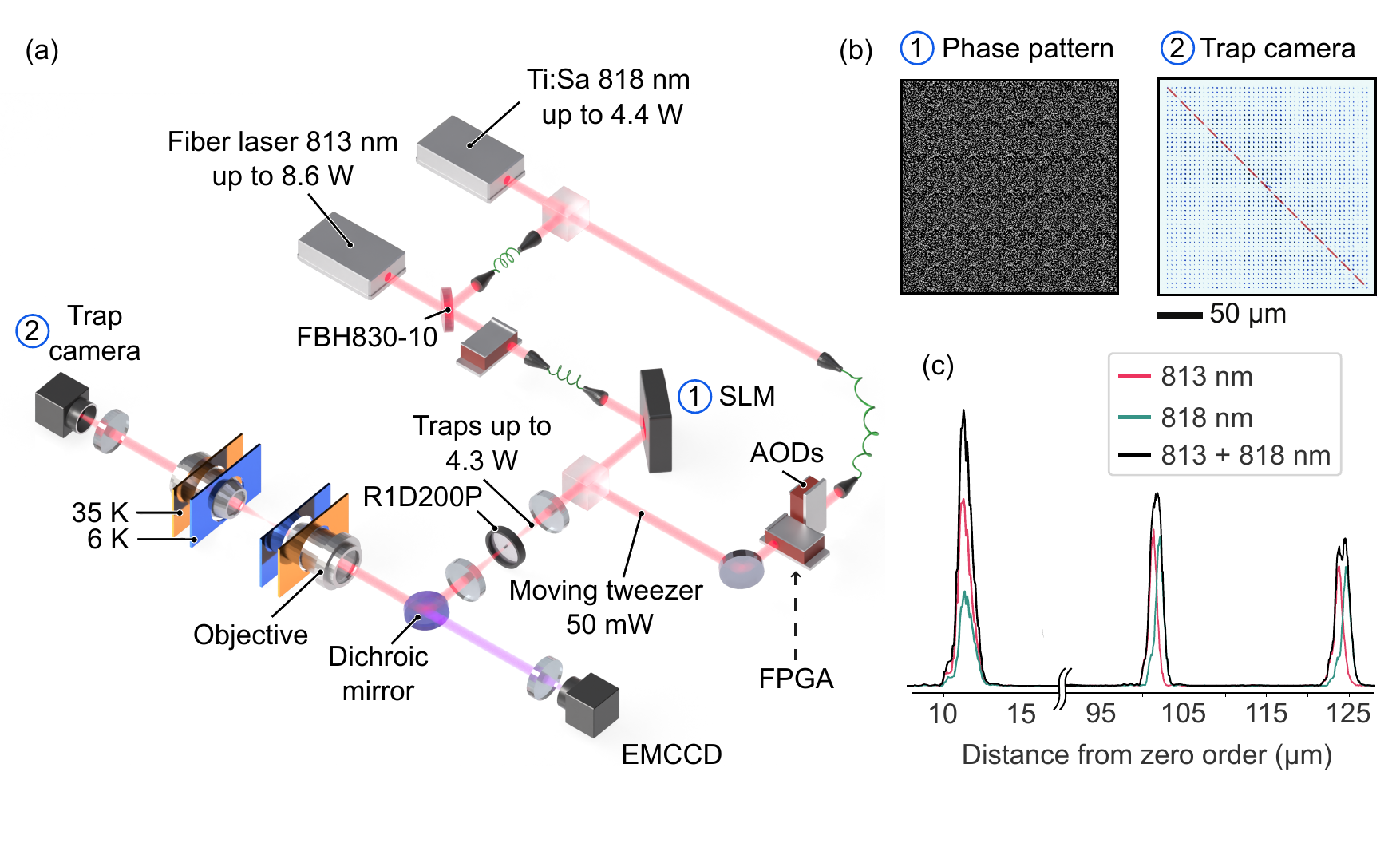}
\caption{(a) Layout of the trapping and imaging optics. An 813-nm fiber laser with an output of up to 8.6\,W is combined with a Ti:Sa laser at 818~nm with up to 4.4\,W using a bandpass interference filter (Thorlabs FBH830-10). Taking into account diffraction efficiencies, we manage to obtain a power of up to 4.3~W after the SLM. Part of the power of the Ti:Sa laser  at 818~nm is used for the moving tweezers, whose position is given by the acousto-optic deflectors driven by an FPGA (see text). An obstruction target (Thorlabs R1D200P) is used to block out most of the residual zeroth order. The beam is then focused by the  objectives in the center of the science chamber. The trap camera is used to diagnose the tweezers array. (b) An example of a phase pattern displayed on the SLM and the corresponding trap pattern created for a $45\times45$ square array of optical tweezers. (c) Variation of the intensity profile of the traps along the direction of the red line in (b), as imaged by the trap camera (by taking three images, with both lasers on, only the 818~nm laser, or only the 813~nm laser).  The combination of both wavelengths lead to a splitting of the trapping potentials at large diffraction angles (note the break in the horizontal axis).}
\label{fig3}
\end{figure*}

Figure ~\ref{fig2}~(c,d) provides more detail of how the objectives are incorporated with the two shields which are used for cryo pumping. The single steel objective mount contains 7-mm diameter holes for the horizontal MOT beams, with baffles to prevent stray light; a total of four additional 7-mm  pairs of holes in the shields are used for the  vertical MOT beams and for observing the MOT fluorescence~\footnote{In hindsight, adding thin glass windows on these openings would have been a better design in order to improve the quality of vacuum; this will be addressed in future work.}. To tackle the lack of optical access, the objective mount is also equipped with a silver-coated in-vacuo mirror that allows us to image the magneto-optical trap. For future studies on Rydberg physics, electrodes are placed in an octupolar configuration within the $6\,\mathrm{K}$ shields, making it possible to compensate for stray electric fields. Figure~\ref{fig2}~(e) shows one of the two elongated superconducting (SC) magnetic-field coils enclosed within the $6\,\mathrm{K}$ shield for the realization of the magneto-optical trap (MOT). In operation, the coils produce a small gradient of a few $\mathrm{G/cm}$ for MOT loading. Additionally, both shields contain long slits to counteract the development of strong eddy currents when the magnetic field is varied. 

While this architecture theoretically provides many advantages over the previous design, one of our main worries was that the shields would not be able to reach the desired temperatures as they are very close to the objectives at $300\,\mathrm{K}$, giving a significant heat load due to black-body radiation at $300\,\mathrm{K}$. In spite of this, the shields can be routinely cooled down to $35\,\mathrm{K}$ and $6\,\mathrm{K}$ respectively within 12 hours [Fig.~\ref{fig2}~(g)]. With the superconducting transition temperature of the MOT coils being around  $10 \,\mathrm{K}$, we can thus operate them without any dissipation. 

A second major challenge to scaling the tweezers array platform is the required laser power. At about 2\,mW per tweezers for a depth of $\sim 1\,\mathrm{mK}$, we need a total of more than $4\,\mathrm{W}$ to create the large pattern shown in Fig.~\ref{fig1}. Moreover, this power budget should take into account the losses in the optical path, where the major contribution comes from the finite efficiency of the optical tweezers generation, which relies on the use of a Spatial Light Modulator (SLM). To ensure sufficient power, for this work our approach consists in combining two lasers of different wavelengths. The wavelengths are chosen to be far enough to be combined on a dichroic mirror, but close enough to avoid excessive trap deformation by the holographic trap generation setup. In practice, we use a Ti:Sa laser (MSquared Solstis) emitting up to $4.4\,\mathrm{W}$ of $818\,\mathrm{nm}$ light, and a fiber laser (Precilaser, model FL-SF-813-10-CW) generating $8.6\,\mathrm{W}$ at $813\,\mathrm{nm}$.

Figure~\ref{fig3}(a) illustrates schematically how we combine the two laser sources. The laser beams are overlapped on an interference filter (FBH830-10 from Thorlabs) that we use as a dichroic mirror by carefully choosing the angle of incidence for the two beams. The combined beam is then fiber-coupled to the optical bench where we generate the traps, thus ensuring co-propagation of the two beams and equal polarization. The array of optical tweezers is created from this beam using a computer-generated hologram displayed on a liquid crystal SLM (Hamamatsu X15213-02)~\cite{nogrette2014}. To compute the phase mask appropriate for a given tweezers array we use a weighted Gerchberg-Saxton algorithm running on a dedicated GPU to  decrease calculation time.

Since the trapping light is bichromatic, the trapping potential results from the incoherent addition of the intensities of the respective diffraction patterns of the two spectral components at $813$ and $818\,\mathrm{nm}$, for which the diffraction angles, for a given phase mask, are slightly different. This means that traps close to the zeroth order are obtained by adding two Gaussian traps that are almost at the same position, while at the periphery of our largest arrays, the two contributions are slightly displaced, resulting in a broader optical tweezers, as can be seen in Fig.~\ref{fig3}(c). For even larger arrays, the overlap between the two sets of traps would vanish at the periphery, and in order to use two different wavelengths for trapping, one should use two distinct SLMs, as done for instance in~\cite{manetsch2024tweezer}.

At large diffraction angles, the SLM diffraction efficiency decreases significantly. Thus, shifting the entire array away from the SLM zeroth-order beam using a blazed grating, as done in our previous works~\cite{nogrette2014,barredo2016atom,schymik2022situ}, becomes inefficient. We estimate an 18\% gain in power in the first order by choosing to center the traps around the zeroth order rather than adding a $100\,\mathrm{\mu m}$ shift.  We place an obstruction target (R1D200P from Thorlabs) in a plane conjugated with the trapping plane to remove most of the contribution of zeroth order that otherwise affects the traps next to it. Still, to further improve the quality of the array, we do not generate traps close to the center of the array, where residual light from the zeroth order creates some speckle; this is the reason behind the presence of the small, diamond-shaped empty region in the center of the array shown in Fig.~\ref{fig1}(b). With this configuration, we reach around $60\%$ of the incoming power in the first diffraction order, giving us up to 4.3\,W on the atoms, enough for the generation of $\sim2000$ trapping sites. With this amount of power sent onto the objectives, we measure a moderate increase ($\sim0.5\,\mathrm{K}$) of the temperature of the inner shield.

We now study the loading of single atoms in these large arrays. The  laser cooling and trapping sequence is as follows. The atomic beam from the 2D MOT directly loads a 3D MOT in the middle of the two objectives. The MOT uses three pairs of retro-reflected $780\,\mathrm{nm}$ laser beams, with a $1/e^2$ radius of $2\,\mathrm{mm}$ each. The MOT beams have a power of $1\,\mathrm{mW}$ and are detuned by $\Delta = -3.2\,\Gamma$ from the $F = 2 \rightarrow F' = 3$ cycling transition of $^{87}$Rb. The cooling light is mixed with a repumper, tuned on resonance with the $F = 2 \rightarrow F' = 2$ transition. The optical tweezers array is overlapped with the MOT for a few seconds before we switch off the atomic source.  

Figure\ref{fig1}(c) shows a single-shot image of a 2088-trap array at the end of the loading procedure, with an occupancy of 53\%. To characterize the traps, we measure, using parametric heating, a transverse trapping frequency of about $105\,\mathrm{kHz}$, from which we calculate a trap depth of $U_0/k_\mathrm{B} \simeq 1.0\,\mathrm{mK}$, in good agreement with the expected value for a $2\,\mathrm{mW}$ of trapping power per tweezers of $1/e^2$ radius $1.0\,\mu{\rm m}$ (these values correspond to traps close to the axis, where the two patterns at $813$ and $818\,\mathrm{nm}$ perfectly overlap). Using a release-and-recapture method, we measure an atomic temperature of about $50\,\mu$K.

\begin{figure}[t!]
\centering
\includegraphics[width=80mm]{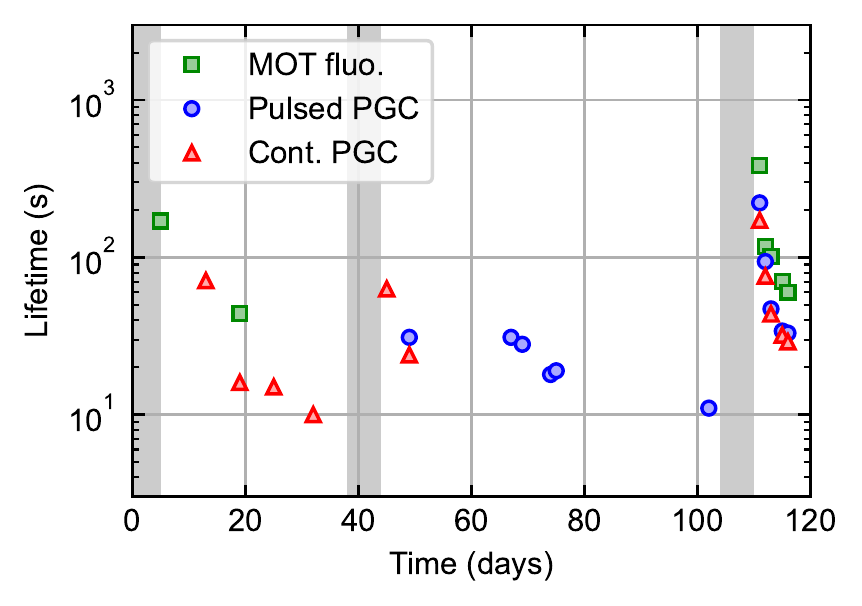}
\caption{Evolution of the atomic lifetime over the course of several months. The gray-shaded areas correspond to periods when the cryostat was warmed up to room temperature. We assess the quality of the vacuum by measuring the MOT decay time (green squares) and the lifetime of the atoms in the optical tweezers in the presence of continuous (red triangles) or pulsed (blue disks) polarization gradient cooling (PGC).   }
\label{fig4}
\end{figure}

As stated above, a disadvantage of our design is the presence of openings in the thermal shields (1-mm wide gaps around the objectives, see Fig.~\ref{fig1}(a),~(iii), and 7-mm diameter holes for the MOT beams, as mentioned above). We anticipated that this could degrade the quality of the vacuum environment experienced by the atoms, since the 300~K vacuum chamber is at a relatively high  pressure (in the $ 10^{-10}$ mbar range), as it cannot be baked because of the presence of the PTR. However, this effect is hard to model \emph{a priori}, so we use direct measurements of the atom trapping lifetime to assess the quality of vacuum. 

We perform three types of measurements, as in~\cite{schymik2021single}, using  (1) the MOT fluorescence decay at long times, or the measured lifetime in a tweezers in the presence of continuous (2) or pulsed (3) polarization gradient cooling (PGC). Figure~\ref{fig4} shows the results of such lifetime measurements, conducted over a period a few months, with three periods where the cryostat was warmed up (gray-shaded areas). The different types of measurements all confirm that the lifetime is initially, just after cool-down, in the range of a couple of hundreds of seconds, with a quick degradation within a few days, and leveling towards values of 10 to 20~s after typically a month. Warming up the cryostat to room temperature, while keeping it under vacuum, regenerating  the NEG pumps, and cooling down again allows to recover long lifetimes.   

Using a residual gas analyzer, we have not been able to detect any  leak in the vacuum system, and we believe that the behavior we observe can be explained as follows.  
Just after cool-down, the inner walls of the 6~K shield are clean, and act as very efficient cryo-pumps. Thus the residual pressure experienced by the atoms is quite low, resulting in long lifetimes. However, due to the openings in the shield, residual gas (mostly ${\rm H}_2$) from the 300~K chamber flows into the cryogenic part, and once more than a monolayer has built up on the walls, cryopumping is gradually replaced by cryo-condensation, which is much less efficient~\cite{baglin2020cryopumping}.
An order-of-magnitude estimate of the time needed to build a ${\rm H}_2$ monolayer, given the geometry of our system and the pressure in the 300~K chamber, gives a typical time of a few days, compatible with the observed slow degradation of the trapping lifetime of the atoms. 

Despite this limited vacuum lifetime, we investigate atom-by atom rearrangement with a moving tweezers as in~\cite{barredo2016atom}.   
An additional fiber-coupled beam is sent from the $818\,\mathrm{nm}$ laser through a pair of orthogonal acousto-optical deflectors (AODs) that we use to steer the moving tweezers. To drive the AODs, we developed a custom solution using two evaluation boards: an AMD ZCU102 ``System on a Chip'' including an ARM CPU and a FPGA, and an Analog Devices AD9154 Digital to Analog Converter. This solution is capable of generating up to four~\footnote{We need only two for the experiments reported here.} fully independent radiofrequency signals with arbitrary amplitude, and frequency/phase modulation on a carrier that can reach up to $400\,\mathrm{MHz}$. All the modulations are calculated in real time with a $4\,\mathrm{ns}$ resolution, well beyond the bandwidth required for rearrangement.

An example of a result of such a rearrangement procedure (with two successive rearrangement cycles~\cite{Schymik2020}) is shown in Fig.~\ref{fig1}(d). The target array, in the shape of an annulus with external (internal) radius of $66\,\mu{\rm m}$ ($25\,\mu{\rm m}$), contains 828 traps, and is filled with 95\% occupancy. The rearrangement requires $\sim800$ moves (including the ones needed to dump unused atoms) and such a cycle lasts about 1.65~s (calculation of moves $\sim150$ ms, transfer of moves to the FPGA $\sim500$ ms, actual execution of moves slightly below one second~\footnote{An individual move consists of a linear ramp up of the moving tweezers intensity in $300\,\mu$s, a linear ramp in frequency to move the atom (with a velocity of $10\,\mu{\rm m}/{\rm ms})$, and a linear ramp down of intensity in $300\,\mu$s at the target trap position. Our FPGA approach allows for immediate improvements on these protocols, such as the use of smooth, sigmoidal-shaped curves for the position of the moving tweezers, and of optimized, adiabatic ramps for picking up and releasing the atoms. At the time of writing, these changes are being implemented on the setup. The generation of  up to 16 multiple tweezers  for parallel rearrangement is also compatible with the hardware.}). Losses due to vacuum-limited lifetime (measured to be about 100~s at the time when the data was taken) during the two rearrangement cycles  thus account for the major part of the observed defects, and improvements in the vaucum quality would directly translate into better assembling efficiencies. 

In conclusion, the setup described in this work successfully combines large field-of view microscope objectives and a cryostat, making it possible to realize, in an environment at 6~K, tweezers arrays at the scale of 2000 traps, and to rearrange atomic registers at the 1000-atom scale. However, the quality of the vacuum remains unsatisfactory so far, as collisions with background gas limit the trapping lifetime at the scale of a few 10~s, which strongly limits the  occupancy of the rearranged arrays. In future work, several directions can  be explored to tackle this issue, such as the use of activated charcoal to enhance cryopumping, and the inclusion of thin glass windows to occlude the 6-K shield openings. Combined with improvements in the rearrangement procedure, in particular using multiple tweezers in parallel as in~\cite{endres2016atom}, obtaining almost defect-free atom arrays at the 1000-atom scale in a cryogenic environment is within reach in the near future. Finally, the current setup, with the presence of electrodes, ITO coating, and appropriate anti-reflection coatings on viewports, is fully compatible with the requirements for controlled Rydberg excitation, opening exciting prospects for quantum science and technology with very large Rydberg arrays~\cite{kaufman2021quantum}. 
 
\begin{acknowledgments}
We thank K.-N. Schymik for insightful discussions, C.~Montmeyran for useful comments on the manuscript, and the whole PASQAL team for invaluable support, especially the electronics, optics and optomechanics teams. AB and TL acknowledge support from the Horizon Europe programme HORIZON-CL4-2022-QUANTUM-02-SGA via the project 101113690 (PASQuanS2.1) and by the Agence Nationale de la Recherche (ANR-22-PETQ-0004 France 2030, project QuBitAF). 
\end{acknowledgments}

\end{document}